\documentclass[11pt]{article}
\usepackage{moriond}

\def\be{\begin{equation}}
\def\ee{\end{equation}}
\def\bea{\begin{eqnarray}}
\def\eea{\end{eqnarray}}

\newcommand{\Photo}{\includegraphics[height=35mm]{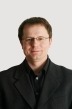}}

\newcommand{\ep}{\mbox{$e^{+}$}}
\newcommand{\el}{\mbox{$e^{-}$}}
\newcommand{\pio}{\mbox{$\pi^{0}$}}
\newcommand{\pim}{\mbox{$\pi^{-}$}}
\newcommand{\pip}{\mbox{$\pi^{+}$}}

\newcommand{\phietaee}{\mbox{$\phi \to \eta \ep\el$}}

\begin{document}
\vspace*{4cm}
\title{RECENT RESULTS ON HADRON PHYSICS AT KLOE}

\author{ 
P. Moskal on behalf of the KLOE and KLOE--2 Collaborations\footnote[1]{The KLOE--2 Collaboration:
D.~Babusci, D.~Badoni, I.~Balwierz-Pytko, G.~Bencivenni, C.~Bini, C.~Bloise, F.~Bossi, P.~Branchini, 
A.~Budano, L.~Caldeira~Balkest\aa hl, G.~Capon, F.~Ceradini, P.~Ciambrone, F.~Curciarello, E.~Czerwi\'nski, E.~Dan\`e,
V.~De~Leo, E.~De~Lucia, G.~De~Robertis, A.~De~Santis, A.~Di~Domenico, C.~Di~Donato, R.~Di~Salvo, D.~Domenici,
O.~Erriquez, G.~Fanizzi, A.~Fantini, G.~Felici, S.~Fiore, P.~Franzini, A.~Gajos, P.~Gauzzi, G.~Giardina,
S.~Giovannella, E.~Graziani, F.~Happacher, L.~Heijkenskj\"old, B.~H\"oistad,
L.~Iafolla, M.~Jacewicz, T.~Johansson, K.~Kacprzak, A.~Kupsc, J.~Lee-Franzini, B.~Leverington, F.~Loddo,
S.~Loffredo, G.~Mandaglio, M.~Martemianov, M.~Martini, M.~Mascolo, R.~Messi, S.~Miscetti, G.~Morello,
D.~Moricciani, P.~Moskal, F.~Nguyen, A. Palladino, A.~Passeri, V.~Patera, I.~Prado~Longhi, A.~Ranieri, C.~F.~Redmer,
P.~Santangelo, I.~Sarra, M.~Schioppa, B.~Sciascia, M.~Silarski, C.~Taccini, L.~Tortora, G.~Venanzoni,
W.~Wi\'slicki, M.~Wolke, J.~Zdebik
}}

\address{Department of Physics, Jagiellonian University, Reymonta 4, Poland}

\maketitle\abstracts{
One of the basic motivations of the KLOE and KLOE-2 collaborations
is the test of fundamental symmetries and the search for phenomena beyond the Standard Model
via the hadronic and leptonic decays of ground-state mesons and via their production in 
the fusion of virtual gamma quanta
exchanged between colliding electrons and positrons.
This contribution includes brief description of results of recent analysis of the KLOE data
aimed at (i) the search for the dark matter boson, (ii) determination of the hadronic and light-by-light contributions to the g-2 muon anomaly
and (iii) tests of QCD  anomalies. 
}

\section{Introduction}  
The KLOE detector 
consists of a $\sim$~3.5~m long cylindrical drift chamber with a diameter of about 4~m
surrounded by the sampling electromagnetic calorimeter~\cite{kloe,EMCkloe,DCkloe}.
Both these detectors
are immersed in the axial 
magnetic field
($\sim0.5$~T) 
provided by the superconducting solenoid.
The detector surrounds the crossing region of the positron and electron beams
circulating in the rings of the DA$\Phi$NE collider~\cite{flavorofKLOE}.

Results presented in this contribution have been obtained using the data sample 
collected by the KLOE collaboration. 
Search for the U boson, 
and studies of the box anomaly  
and
$e^+e^- \to \pi^+\pi^-$ process 
were based on the data taken at the center-of-mass energy of $\sqrt{s}$~=~1.02 GeV corresponding 
to the mass of the $\phi$ meson, whereas the studies of the   
$\gamma^*\gamma^* \to \eta$ process 
were based on the data sample 
taken at the center-of-mass energy of $\sqrt{s}$ = 1 GeV,
 where background from $\phi$ meson
decay is suppressed.

\section{Search for the dark matter boson}
\begin{figure}[!h]
  \begin{center}
    \includegraphics[width=0.44\textwidth]{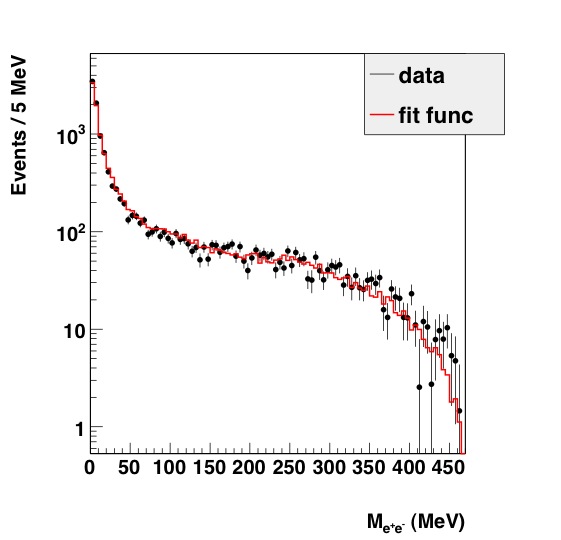}
    \includegraphics[width=0.44\textwidth]{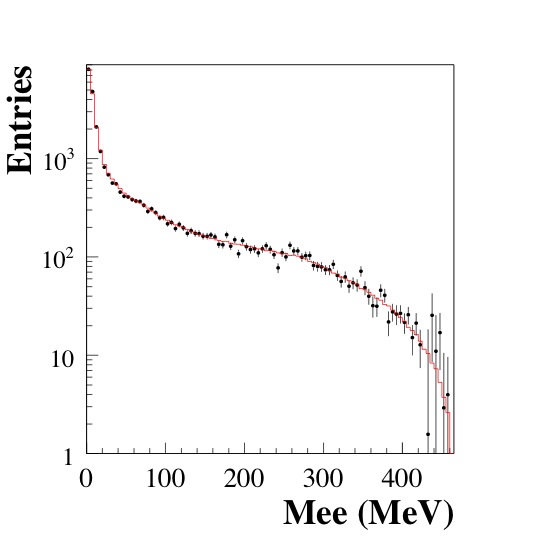}
     \caption{
     ${e^+e^-}$ invariant mass spectrum for $\phietaee$ decay with $\eta\to\pip\pim\pio$ (left) and 
     with $\eta\to3\pio$ (right). Solid lines indicate result of the fit  
performed assuming the Vector Meson Dominance expectations for the 
$\phi\eta\gamma^*$ transition form factor. 
  \label{fit_zdebik_sarra}
     }
 \end{center}
\end{figure}

There are many astrophysical obsrevations indicating existence of dark matter.
For example:
an excess of the $e^+ e^-$ annihilation $\gamma$ quanta
from the galactic center observed by
the INTEGRAL satellite~\cite{Jean:2003ci}, the excess in the
cosmic ray positrons reported by PAMELA~\cite{Adriani:2008zr}, the total
electron and positron flux measured by ATIC~\cite{Chang:2008zzr}, Fermi~\cite{Abdo:2009zk},
and HESS \cite{Aharonian:2009ah},
and the annual modulation of the DAMA/LIBRA signal~\cite{Bernabei:2008yi}.
The origin of this kind of enhanced stream of radiation
may be explained assuming~\cite{kloe2amelino}
that positrons are created in an annihilation of the dark matter particles into $e^+ e^-$ pairs,
and that this process is mediated by the
U boson with mass in the GeV
scale. 
The existence of such U boson can manifest itself 
as a maximum in the invariant mass distribution of $e^+ e^-$ pairs originating
from the radiative decays as e.g. 
$\phi\to\eta\ep\el$. In this case a light dark-force mediator (U boson) may contribute to this process 
via following decay chain:   
$\phi\to\eta\gamma^*\to\eta\,U\to\eta\gamma^*\to\eta\ep\el$. 
In Figure~\ref{fit_zdebik_sarra} we present results of the analysis
of data sample of 
1.7 fb$^{-1}$ where
no structures are observed in the \ep\el\ invariant mass distribution 
over the background.
Therefore, we set only
an upper limit at 90\% C.L.\  
on the ratio between the $U$ boson coupling constant and the fine 
structure constant of 
$\alpha'/\alpha < 1.7 \times 10^{-5}$ for $30<M_{U}<400$~MeV and
$\alpha'/\alpha \leq 8 \times 10^{-6}$ for the sub-region 
$50<M_{U}<210$ MeV~\cite{Uboson}.

\section{KLOE contribution to the determination of the g-2 anomaly}

Comparison of measured and calculated value of the muon magnetic moment anomaly
$a_{\mu} = (g_{\mu}-2)/2$
constitutes one of the most precise
test of the Standard Model,  since
$a_{\mu}$ was measured with the
precision of
0.5 ppm~\cite{Bennett:2006fi},
and
FNAL experiment~\cite{Carey:2009zz}
plans to improve this accuaracy to 0.14 ppm in the near future.
The predictions of the value of 
$a_{\mu}$ 
based on the SM are however limited by the accuracy
of the determination of the hadronic contributions
which in 70\% originates from the two pion contribution due to the $\gamma^*\to\pi^+\pi^-$ process.
Therefore, we have conducted the independent measurement of the
$e^+e^-\to \pi^+\pi^-(\gamma)$ cross section below 1 GeV, which is
particularly important to test the Standard Model calculation for
the (g-2) of the muon, where a long standing 3 sigma discrepancy
is observed. 

\begin{figure}[!h]
  \begin{center}
    \includegraphics[width=0.52\textwidth]{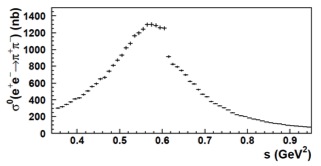}
     \caption{
       The bare cross section from the $\pi^+\pi^-\gamma/\mu^+\mu^-\gamma$ 
       ratio~\protect\cite{KLOEg-2}.
  \label{wykres15}
     }
\end{center}
\end{figure}

We have determined the ratio of $\sigma(e^+e^-\rightarrow\pi^+\pi^-\gamma)/\sigma(e^+e^-\rightarrow \mu^+\mu^-\gamma)$, 
using a total integrated luminosity of about 240 pb$^{-1}$.
From this ratio we obtain the cross section $\sigma(e^+e^-\rightarrow\pi^+\pi^-)$ shown in Figure~\ref{wykres15}.
From the cross section we determine the pion form factor 
$|F_\pi|^2$ 
and the two-pion contribution to the muon anomaly $a_\mu$ for $0.592<M_{\pi\pi}<0.975$ GeV, 
{\bf $\Delta^{\pi\pi} a_\mu$= $({\rm 385.1\pm1.1_{stat}\pm2.7_{sys+theo})}\times10^{-10}$}. 
This result confirms the current discrepancy between the Standard Model calculation 
and the experimental measurement of the muon anomaly.

It is worth mentioning that 
the previous KLOE measurements 
were normalized to the
luminosity using large angle Bhabha scattering, 
whereas the dereviation of $\sigma(e^+e^-\rightarrow\pi^+\pi^-)$
from the ratio of cross sections caused cancelation of many potential sources of uncertainty as e.g.:
the radiator function, luminosity derivation, vacuum polarization corections, 
and to large extent also acceptance corrections~\cite{KLOEg-2}.
In addition the influence of FSR was minimized by taking into account only small angular range for $\gamma$ quanta.

Another large contribution to the uncertainty of the 
$a_{\mu}$ 
calculations 
originates from the uncertainty in determination  
of pseudoscalar transition form factors which
dominates the precision in determination of hadronic light-by-light contributions. 
Therefore the precise studies of transition form factors
is of importance for the SM predictions of the
anomalous magnetic moment of the muon.

Studies of the
conversion decays give information about the time-like region of the form-factor
with positive $q^2$
equal to the square of the invariant mass of the $l^+ l^-$ pair~\cite{KLOE-TFF1,KLOE-TTF2}.
Information about the space-like region with the negative values of $q^2$ is accessible
via cross section of mesons production in  $\gamma^* \gamma^*$ fusion realized in e.g.
$e^+ e^- \to e^+ e^- \gamma^*\gamma^* \to e^+ e^- \eta$ reaction~\cite{JHEP}. From the measurement
of the cross section of this process we derived the partial width $\Gamma(\eta\to\gamma\gamma)= (520\pm20_{stat}\pm13_{syst})eV$~\cite{JHEP},
which is the most precise measurement to date.

\section{Tests of QCD anomalies}
The $\eta$ meson is particularly suited for studies of QCD anomalies because
all its strong and electromagnetic decays are forbidden in the first order~\cite{nefkens}.
    The most energetically favourable strong decay of $\eta$ into  $2\pi$ is forbidden
    due to P and CP invariance.
    Its decay into $3\pi$ is suppressed by G-parity and isospin invariance~\cite{marcinmgr},
    and it occurs due to the  difference between the mass of $u$ and $d$ quarks.
    The first order electromagnetic decays as $\eta \to \pi^0 \gamma$  or $\eta \to 2\pi^0 \gamma$  break
    charge conjugation invariance and  $\eta \to \pi^+ \pi^- \gamma$
    is also suppressed because charge conjugation
 conservation requires odd (and hence nonzero) angular momentum in the $\pi^+ \pi^-$ system.
    Therefore, this radiative decay at a massless quark limit
    is driven by the QCD box anomaly. 

The ratio  $R_{\eta}=\Gamma(\eta \to \pi^+\pi^-\gamma)/\Gamma(\eta \to \pi^+\pi^-\pi^0)$ 
has been measured by analysing 22 million
$\phi \to \eta \gamma$ decays collected by the KLOE
experiment, corresponding to an
integrated luminosity of 558 pb$^{-1}$. The $\eta \to
\pi^+\pi^-\gamma$ proceeds both via the $\rho$ resonant
contribution, and possibly a non-resonant direct
term connected to the box anomaly. Our result, $R_{\eta}= 0.1856\pm
0.0005_{stat} \pm 0.0028_{syst}$, points out a sizable contribution of
the direct term to the total width~\cite{QCD}.
The di-pion invariant mass for the $\eta \to \pi^+\pi^-\gamma$ decay
could be described in a model-independent approach in terms of a
single free parameter, $\alpha$. The determined value of the
parameter $\alpha$ is equal to 
$(1.32 \pm 0.08_{stat}$$^{+0.10} _{-0.09}$$_{syst}$$\pm 0.02_{theo})$ 
GeV$^{-2}$~\cite{QCD}, and it is in agreement with the result of the WASA collaboration~\cite{WASA}. 

\section{Perspectives}
Taking atvantage of 
a successfuly commissioned~\cite{zubov}
new  electron-positron interaction region of DA$\Phi$NE,
in the near future the data sample will be significantly increased by means of the KLOE-2 detector setup~\cite{kloe2amelino,kloe2_fabio},
which is a successor of KLOE
upgraded with 
new components
in order to improve its tracking and clustering
capabilities as well as in order
to tag $\gamma\gamma$ fusion processes~\cite{LET,HET,IT,QCAL,CCAL}.

\section*{Acknowledgments}
This work was supported in part by the EU 
Hadron Physics Project under contract number RII3-CT-2004-506078; by the European
Commission: 
FP7-INFRASTRUCTURES-2008-1, Grant Agreement No. 283286; by the Polish
National Science Centre through the 
Grants No. 0469/B/H03/2009/37,
2011/01/D/ST2/00748, 2011/03/N/ST2/02652,
2011/03/N/ST2/02641 and by the Foundation for Polish Science through the MPD programme 
and the project HOMING PLUS BIS/2011-4/3.

\end{document}